\documentclass[letterpaper, 10 pt, conference]{ieeeconf}  

\IEEEoverridecommandlockouts                              

 \usepackage[bookmarks=false]{hyperref}

 \usepackage{float}
\usepackage[noadjust]{cite}
\usepackage{relsize}
\usepackage{amsfonts}
\usepackage{mathrsfs}
\usepackage{bbm}
\usepackage{pifont}
\usepackage{amsfonts}
\usepackage{amsmath}

\usepackage{tkz-graph}
\usetikzlibrary{calc}
\tikzset{me/.style={to path={
\pgfextra{%
 \pgfmathsetmacro{\startf}{-(#1-1)/2}
 \pgfmathsetmacro{\endf}{-\startf}
 \pgfmathsetmacro{\stepf}{\startf+1}}
 \ifnum 1=#1 -- (\tikztotarget)  \else
     let \p{mid}=($(\tikztostart)!0.5!(\tikztotarget)$)
         in
\foreach \i in {\startf,\stepf,...,\endf}
    {%
     (\tikztostart) .. controls ($ (\p{mid})!\i*6pt!90:(\tikztotarget) $) .. (\tikztotarget)
      }
      \fi
     \tikztonodes
}}}

 \usepackage[usenames,dvipsnames]{pstricks}
 \usepackage{epsfig}
 \usepackage{pst-grad} 
 \usepackage{pst-plot} 

\usepackage[top=.75in,bottom=.75in,left=.75in,right=.75in]{geometry}

\usepackage[font=normalsize,labelfont=normalsize]{caption}
\usepackage{makecell}
\usepackage{booktabs}

\def\hlinewd#1{%
\noalign{\ifnum0=`}\fi\hrule \@height #1 %
\futurelet\reserved@a\@xhline}

\usepackage{tabularx}
\usepackage{listings}
\usepackage{graphicx}
\usepackage{float}
\usepackage{amssymb}
\usepackage{array}
\usepackage{fancyhdr}
\usepackage{cases}
 \usepackage{supertabular}
\newcommand{\bcap} {\hspace{2pt} \mathlarger{\cap}
\hspace{2pt}}

\newcommand{\scap} {\mathsmaller{\bigcap}}

\usepackage{fancyhdr}

\usepackage{psfrag}
\usepackage{color}
\usepackage{bbding}

\newcommand{\qeda}{\hfill\ensuremath{\blacksquare}}
\newcommand{\qedb}{\hfill\ensuremath{\square}}

\newcommand{\bP}[1]{{\mathbb{P}}\left[{#1}\right]}

\def\centerhack#1{\hbox to 0pt{\hss\footnotesize #1\hss}}
\def\centerhackn#1{\hbox to 0pt{\hss #1\hss}}
\def\dchack#1{\vbox to 0pt{\vss{\hbox to 0pt{\hss#1\hss}}\vss}}

\tikzset{>=latex}

\newtheorem{fact}{Fact}
\newtheorem{lem}{Lemma}
\newtheorem{thm}{Theorem}

\newtheorem{rem}{Remark}

\title{\vspace*{.25in}On the Strengths of
Connectivity and Robustness\\ 
 in General Random Intersection Graphs}

\author{ \parbox{2 in}{\centering Jun Zhao
       \\
        CyLab and Dept.
of ECE\\
        Carnegie Mellon University\\
        Pittsburgh, PA 15213, USA\\
        {\tt\small junzhao@junzhao.info}} \hspace*{ 0.05 in}
        \parbox{2 in}{ \centering Osman Ya\u{g}an\\
       CyLab and Dept.
of ECE \\
        Carnegie Mellon University\\
        Moffett Field, CA 94035, USA\\
        {\tt\small oyagan@andrew.cmu.edu}} \hspace*{ 0.05 in}
        \parbox{2 in}{ \centering Virgil Gligor\\
       CyLab and Dept.
of ECE \\
        Carnegie Mellon University\\
        Pittsburgh, PA 15213, USA\\
        {\tt\small gligor@cmu.edu}}
}

\begin{document}

\makeatletter
\let\old@ps@headings\ps@headings
\let\old@ps@IEEEtitlepagestyle\ps@IEEEtitlepagestyle
\def\confheader#1{%
  \def\ps@headings{%
    \old@ps@headings%
    \def\@oddhead{\strut\hfill#1\hfill\strut}%
    \def\@evenhead{\strut\hfill#1\hfill\strut}%
  }%
  \def\ps@IEEEtitlepagestyle{%
    \old@ps@headings%
    \def\@oddhead{\strut\hfill#1\hfill\strut}%
    \def\@evenhead{\strut\hfill#1\hfill\strut}%
  }%
  \ps@headings%
} \makeatother \lhead{}

\rhead{}

\maketitle

\thispagestyle{fancy} \pagestyle{fancy}

\markboth{a}{b}

\fancyhead[C]{{IEEE Conference on Decision and Control (CDC) 2014}}

\begin{abstract}

Random intersection graphs have received much attention for nearly
two decades, 
 and currently have a wide range of applications ranging from key predistribution in wireless sensor
 networks to modeling social networks. In this paper, we investigate the
strengths of connectivity and robustness in a {\em general random
intersection graph} model. Specifically, we establish sharp asymptotic zero--one laws for 
 $k$-connectivity and $k$-robustness, as well as the
asymptotically exact probability of $k$-connectivity,
 for any positive integer $k$.
The $k$-connectivity property quantifies
how resilient is the connectivity of a graph against node or edge
failures. On the other hand, $k$-robustness measures the effectiveness of {\em local} diffusion strategies (that do not use global graph topology information) in spreading information over the graph
in the presence of misbehaving nodes. 
  In addition to presenting the results under the general random
intersection graph model, we consider two special cases of the
general model, a {\em binomial} random intersection graph and a {\em
uniform} random intersection graph, which both have numerous applications as well.
For these two specialized graphs, our results on asymptotically
exact probabilities of $k$-connectivity and asymptotic zero--one laws for $k$-robustness are also novel in the literature.

\end{abstract}

\begin{keywords}
Connectivity, consensus,
random graph, random intersection graph, random key graph, robustness.
\end{keywords}

\section{Introduction}

\subsection{Graph Models} \label{sec:GraphModel}

Random intersection graphs have been introduced by Singer-Cohen
\cite{RIGThesis} and received considerable attention \cite{Rybarczyk,ZhaoYaganGligor,yagan,ISIT,virgil,ZhaoAllerton,ball2014,mil10,2013arXiv1301.7320B,Models,2013arXiv1301.0466R,zz,r1,ryb3,bloznelis2013,virgillncs} for nearly two decades. In these graphs,
each node is assigned a set
of {\em objects} selected by some random mechanism. An undirected edge exists between any two nodes that
have at least one object in common.
Random intersection graphs have proved useful in modeling and analyzing real-world networks in a wide variety of application areas. Examples include
secure wireless sensor networks  \cite{Rybarczyk,ZhaoYaganGligor,yagan,ISIT,virgil,ZhaoAllerton}, frequency hopping spread spectrum
 \cite{ZhaoAllerton}, spread of epidemics \cite{ball2014,mil10}, and social and information networks \cite{bloznelis2013,mil10,ZhaoYaganGligor} including collaboration networks \cite{bloznelis2013,mil10} and common-interest networks \cite{ZhaoYaganGligor}.
Several classes of random intersection graphs have been analyzed, and results concerning
various graph properties such as clustering \cite{bloznelis2013},
component evolution \cite{2013arXiv1301.7320B,Rybarczyk} and degree
distribution \cite{Models} have been obtained.

The model considered in this paper, hereafter referred to as a {\em
general random intersection graph}, represents a
 generalization \cite{bloznelis2013,Models,Rybarczyk} of  random intersection
graphs. It is defined on a node set $\mathcal {V} = \{v_1,
v_2, \ldots, v_n \}$ as follows. Each node $v_i$ ($i=1,2,\ldots,n$)
is assigned an object set $S_i$ from an object pool $\mathcal {P}$
consisting of $P_n$ distinct objects, where $P_n$ is a function of
$n$. Each object $S_i$ is constructed using the following two-step
procedure: First, the size of $S_i$, $|S_i|$, is determined
according to some probability distribution $\mathcal {D}:\{1,
2,\ldots, P_n\} \to [0,1]$. Of course, we have $\sum_{x = 1}^{P_n}
\mathbb{P}[|S_i| = x] = 1$, with $\mathbb{P}[A]$ denoting the
probability that event $A$ occurs. Next, $S_i$ is formed by
selecting $|S_i|$ distinct objects uniformly at random from the
object pool $\mathcal {P}$. In other words, conditioning on $|S_i| =
s_i$, set $S_i$ is chosen uniformly among all $s_i$-size subsets of
$\mathcal {P}$. This process is repeated independently for all
object sets $S_1, \ldots, S_n$. Finally, an undirected edge is
assigned between two nodes if and only if their corresponding object
sets have at least one object in common; namely, distinct nodes
$v_i$ and $v_j$ have an edge in between if and only if $S_i \bcap
S_j \neq \emptyset$. The graph defined through this adjacency notion
is denoted by $G(n,P_n,\mathcal {D})$.

A specific case of the general model $G(n,P_n,\mathcal {D})$,
known as the \emph{binomial} random intersection graph, has
been widely explored to date [9]--[14].
Under this
model, each object set $S_i$ is constructed by a Bernoulli-like
mechanism; i.e., by adding each object to $S_i$ independently with
probability $p_n$. Like integer $P_n$, probability $p_n$ is also a function of $n$. The term \lq\lq
binomial" accounts for the fact that $|S_i|$ now follows a binomial
distribution with $P_n$ as the number of trials and $p_n$ as the
success probability in each trial. We denote the binomial random
intersection graph by $G_b(n,P_n,p_n)$, where subscript ``b'' stands for ``binomial''.


Another well-known special case of the general model $G(n,P_n,\mathcal
{D})$ is the \emph{uniform} random intersection graph
\cite{r1,yagan,ryb3,virgil,virgillncs,ISIT}. Under the uniform model, the
probability distribution $\mathcal {D}$ concentrates on a single
integer $K_n$, where $1\leq K_n \leq P_n$; i.e., for each node
$v_i$, the object set size $|S_i|$ equals $K_n$ with probability
$1$. $P_n$ and $K_n$ are both integer functions of $n$. We denote by $G_u(n,P_n,K_n)$ the uniform random intersection
graph, with ``u'' meaning ``uniform''.

A concrete example for the application of random intersection graphs can be given in the context of secure wireless sensor networks. As explained in detail in numerous other places \cite{Rybarczyk,ZhaoYaganGligor,yagan,ISIT,ZhaoAllerton,ball2014,mil10,2013arXiv1301.7320B}, the uniform random intersection graph model $G_u(n,P_n,K_n)$
is induced naturally by the Eschenauer--Gligor (EG) random key predistribution
scheme \cite{virgil}, which is a typical solution to ensure secure
communications in wireless sensor networks. In particular, let the set of $n$ nodes in graph
$G_u(n,P_n,K_n)$ stand for the $n$ sensors in the wireless network. Also, let the object pool $\mathcal{P}$ (with size $P_n$)
represent the set of cryptographic keys available to the network and let $K_n$ be the number of keys assigned to each sensor
(selected uniformly at random from the key pool $\mathcal{P}$). Then, the edges in $G_u(n,P_n,K_n)$ represent pairs of sensors that share at least one cryptographic key and thus that can
\emph{securely} communicate over existing wireless links in the EG scheme. In the above application, objects that nodes have are cryptographic keys, so uniform random intersection graphs are also referred to as random key graphs \cite{virgillncs,yagan,ZhaoAllerton}.

In the secure sensor network area, the general random intersection graph model in this paper captures the differences that may exist among the number of keys possessed
by each sensor. This may occur for various reasons that include (a) the assigned numbers of
keys on sensors may vary prior to deployment given the
heterogeneity in available sensor memory
\cite{Rybarczyk}; (b) the number of keys available to
a sensor may decrease after deployment due to revocation of compromised keys  \cite{ZhaoYaganGligor}; and (c) the number of keys
on a sensor may increase due to the path key establishment phase of the EG scheme \cite{virgil}, where new
path keys are generated and distributed to participating sensors.

\subsection{($k$-)Connectivity and ($k$-)Robustness} \label{sec:ConnectivityandRobustness}
We now introduce the graph properties that we are interested in.
First, a graph is connected if there exists at least a path of edges between any two nodes \cite{citeulike:4012374}.
A graph is said to be $k$-connected if each pair of nodes has at least $k$
internally node-disjoint path(s) in between \cite{zz}; equivalently, a graph is $k$-connected if
it can not be made disconnected by deleting at most $(k-1)$ nodes or edges.\footnote{As in much other work \cite{2013arXiv1301.0466R,zz,FJYGISIT2014,mobihocQ1,ZhaoISIT2014,ANALCO}, $k$-connectivity in this paper means $k$-vertex-connectivity in graph theory \cite{citeulike:4012374,SODAMIT,ANALCO}. Yet,  results on $k$-edge-connectivity similar to those in Theorems 1--3 of Section \ref{sec:kconasy} are shown to hold as well in the full version \cite{fullver}.}
In this manner, $k$-connectivity quantifies the
resiliency of graph connectivity against node or edge failures. In addition, it enables multi-path routing, and is also
useful to achieve consensus in the graph \cite{ZhaoYaganGligor}. In particular, to achieve consensus in the presence of $m$ adversarial nodes in a large-scale graph (with node size greater than $3m$), a necessary and sufficient condition is that the graph is $(2m+1)$-connected  \cite{Dolev:1981:BGS:891722}.

Many algorithms have been proposed to achieve consensus \cite{6425841,add-ref-1,add-ref-2,4738888,5409530,6249754,6760870} in graphs with sufficient connectivity. However, these algorithms typically assume that nodes have full knowledge of the graph topology, which is impractical in some cases \cite{6425841}. To this end, Zhang and
Sundaram \cite{6425841} introduce the notion of {\em graph robustness}. They show that when nodes are limited to local information instead of the global graph topology, consensus can be reached in a sufficiently robust graph in the presence of adversarial/misbehaving nodes, but not in a sufficiently connected and insufficiently robust graph. Therefore, graph robustness quantifies the effectiveness and resiliency of local-information-based consensus algorithms in the presence of adversarial/misbehaving nodes.
Robustness is an important property with broad relevance in graph processes beyond consensus; e.g., robustness plays a key role in information cascades and contagion processes \cite{6425841}. It is worth noting that robustness is a stronger property than connectivity in the sense that any $k$-robust graph is also $k$-connected, whereas a $k$-connected graph is not necessarily $k$-robust \cite{6425841}.


Formally, a graph with a node set $\mathcal {V}$ is $k$-robust
 if at least one of (a) and (b) below hold for any
non-empty and strict subset $T$ of $\mathcal {V}$: (a) there exists
at least a node $v_a \in T$ such that
  $v_a$ has no less than $k$ neighbors inside $\mathcal {V}\setminus
  T$; and (b) there exists at least a node $v_b \in \mathcal {V}\setminus T$ such that
  $v_b$ has no less than $k$ neighbors inside $T$.
    %

\vspace{-3pt}

\subsection{Contributions and Organization}\vspace{-3pt}

With various applications of random intersection graphs, and $k$-connectivity and $k$-robustness graph properties in mind, a natural question to ask is whether random intersection graphs are $k$-connected or $k$-robust  under certain conditions? Our paper answers this question. We summarize our contributions as follows:

\begin{enumerate}
  \item [i)] We derive sharp zero--one laws and asymptotically
  exact probabilities for 
$k$-connectivity in general random intersection graphs.
  \item [ii)] We establish sharp zero--one laws for 
$k$-robustness in general random intersection graphs.
  \item [iii)] For the two specific instances of
the general graph model, a binomial random intersection graph and a
uniform random intersection graph, we provide the first 
results on the asymptotically exact probabilities of $k$-connectivity and zero--one
laws for $k$-robustness.  
\end{enumerate}

The rest of the paper is organized as follows. Section \ref{sec:main:res} presents the main results as
Theorems  \ref{thm:grig}--\ref{thm:urig:rb}. Then, we introduce some auxiliary facts and lemmas in
Section \ref{sec:factlem}, before establishing the main results in
Sections \ref{sec:thmprf:kcon} and \ref{sec:thmprf:krb}. Section
\ref{sec:prf:fact:lem} details the proofs of the lemmas. We provide
numerical experiments in Section \ref{sec:expe}. Section
\ref{related} reviews related work; and Section \ref{sec:Conclusion}
concludes the paper.\vspace{-3pt}

\section{The Results} \label{sec:main:res}

Our main results are presented in Theorems \ref{thm:grig}--\ref{thm:urig:rb} below. We defer the proofs of all
theorems to Sections \ref{sec:thmprf:kcon} and \ref{sec:thmprf:krb}.
Throughout the paper, $k$ is a positive integer and does not scale
with $n$; and $e$ is the base of the natural logarithm function,
$\ln$. All limits are understood with $n\to \infty$. We use the
standard Landau asymptotic notation $o(\cdot), O(\cdot),
\omega(\cdot), \Omega(\cdot),\Theta(\cdot)$ and $ \sim$; in
particular, for two positive functions $f(n)$ and $g(n)$, the relation $f(n) \sim
g(n)$ signifies $\lim_{n \to
  \infty} {f(n)}/{g(n)}=1$. For a random variable $X$, the\vspace{1pt} terms $\mathbb{E}[X]$
and $\textrm{Var}[X]$ stand for its expected value and variance,
respectively.

\subsection{Zero--One Laws and Exact Probabilities for Asymptotic
$k$-Connectivity} \label{sec:kconasy}

We provide zero--one laws and exact probabilities for asymptotic
$k$-connectivity in different graphs below.

\subsubsection{$k$-Connectivity in General Random Intersection
Graphs}

Theorem 1 below presents a zero--one law and the
exact  probability for asymptotic $k$-connectivity in a general
random intersection graph.

\begin{thm} \label{thm:grig} Consider a general random intersection graph
$G(n,P_n,\mathcal {D})$. Let $X$ be a random variable following
probability distribution $\mathcal {D}$. With a sequence $\alpha_n$
for all $n $ defined through\vspace{-3pt}
\begin{align}
 \frac{\big\{\mathbb{E}[X]\big\}^2}{P_n} & =
 \frac{\ln  n + {(k-1)} \ln \ln n + {\alpha_n}}{n}, \vspace{-3pt}
 \label{thm:grig:pe}
\end{align}
%
%
 if $\mathbb{E}[X] = \Omega\big(\sqrt{\ln n}\hspace{2pt}\big)$, $\textrm{Var}[X] =
o\mathlarger{\mathlarger{\big(}}\frac{\{\mathbb{E}[X]\}^2}{ n(\ln
n)^2 }\mathlarger{\mathlarger{\big)}}$ and $|\alpha_n| = o(\ln n)$,
then
\begin{align}\vspace{-3pt}
 & \lim_{n \to \infty}\mathbb{P} \big[\hspace{2pt}\textrm{Graph }G(n,P_n,\mathcal {D})\textrm{
is $k$-connected}.\hspace{2pt}\big]\vspace{-3pt} \nonumber \\ &
\quad =
\begin{cases} 0, &\textrm{ if $\lim_{n \to \infty}{\alpha_n}
=-\infty$}, \\  1, &\textrm{ if $\lim_{n \to \infty}{\alpha_n}
=\infty$,}  \\ e^{- \frac{e^{-\alpha ^*}}{(k-1)!}},
 &\textrm{ if $\lim_{n \to \infty}{\alpha_n}
=\alpha ^* \in (-\infty, \infty)$.} \end{cases}\vspace{-3pt}
\nonumber
 \end{align}

\end{thm}\qedb

\subsubsection{$k$-Connectivity in Binomial
 Random Intersection Graphs}
\indent  Theorem \ref{thm:rig} below presents a zero--one law and
the exact probability for asymptotic $k$-connectivity in a binomial
random intersection graph.

\begin{thm} \label{thm:rig} For a binomial random intersection graph
$G_b(n,P_n,p_n)$, with a sequence $\alpha_n$ for all $n $ defined
through
\begin{align}
  {p_n}^2 P_n & =
 \frac{\ln  n + {(k-1)} \ln \ln n + {\alpha_n}}{n},  \label{thm:rig:pe}
\end{align}
if $P_n = \omega \big(n(\ln n)^5\big)$ and $|\alpha_n| = o(\ln n)$,
then
\begin{align}
 & \lim_{n \to \infty}\mathbb{P}
\big[\hspace{2pt}\textrm{Graph }G_b(n,P_n,p_n)\textrm{
is $k$-connected}.\hspace{2pt}\big] \nonumber \\
& \quad =
\begin{cases} 0, &\textrm{ if $\lim_{n \to \infty}{\alpha_n}
=-\infty$}, \\  1, &\textrm{ if $\lim_{n \to \infty}{\alpha_n}
=\infty$,} \\ e^{- \frac{e^{-\alpha ^*}}{(k-1)!}},
 &\textrm{ if $\lim_{n \to \infty}{\alpha_n}
=\alpha ^* \in (-\infty, \infty)$.} \end{cases}\nonumber
 \end{align}

 \end{thm}\qedb


 \begin{rem} \label{rm}

As we will explain in Section \ref{pfrig} within the proof of
Theorem \ref{thm:rig}, for the zero--one law, the condition $P_n =
\omega \big(n(\ln n)^5\big)$ can be weakened as $P_n = \Omega
\big(n(\ln n)^5\big)$, while we enforce $P_n = \omega \big(n(\ln
n)^5\big)$ for the asymptotically exact probability result. \qedb
 \end{rem}

\subsubsection{$k$-Connectivity in
 Uniform Random Intersection Graphs}

\indent Theorem \ref{thm:urig} below presents a zero--one law and the exact
probability for asymptotic $k$-connectivity in a uniform random
intersection graph.

\begin{thm} \label{thm:urig} For a uniform random intersection graph
$G_u(n,P_n,K_n)$, with a sequence $\alpha_n$ for all $n $ defined through
\begin{align}
 \frac{{K_n}^2}{P_n} & = \frac{\ln  n + {(k-1)} \ln \ln n +
 {\alpha_n}}{n},  \label{thm:urig:pe}
\end{align}
%
%
%
 if $K_n = \Omega \big(\sqrt{\ln n}\hspace{2pt}\big)$ and $|\alpha_n| = o(\ln n)$, then
\begin{align}
 & \lim_{n \to \infty}\mathbb{P} \left[\hspace{2pt}\textrm{Graph }G_u(n,P_n,K_n)\textrm{
is $k$-connected}.\hspace{2pt}\right] \nonumber \\
& \quad =
\begin{cases} 0, &\textrm{ if $\lim_{n \to \infty}{\alpha_n}
=-\infty$}, \\  1, &\textrm{ if $\lim_{n \to \infty}{\alpha_n}
=\infty$,} \\ e^{- \frac{e^{-\alpha ^*}}{(k-1)!}},
 &\textrm{ if $\lim_{n \to \infty}{\alpha_n}
=\alpha ^* \in (-\infty, \infty)$.} \end{cases} \nonumber
 \end{align}

\end{thm}\qedb

\subsection{Zero--One Laws for Asymptotic $k$-Robustness}
\label{sec:main:res:rb}

We provide zero--one laws for asymptotic $k$-robustness in different
graphs below.

\subsubsection{$k$-Robustness in General Random Intersection
Graphs}

Theorem \ref{thm:grig:rb} as follows gives a zero--one law for
asymptotic $k$-robustness in a general random intersection graph.

\begin{thm} \label{thm:grig:rb} Consider a general random intersection graph
$G(n,P_n,\mathcal {D})$. Let $X$ be a random variable following
probability distribution $\mathcal {D}$. With a sequence $\alpha_n$
for all $n $ defined through
\begin{align}
 \frac{\big\{\mathbb{E}[X]\big\}^2}{P_n} & =
 \frac{\ln  n + {(k-1)} \ln \ln n + {\alpha_n}}{n},
 \label{thm:grig:pe:rb}
\end{align}
if $\mathbb{E}[X] = \Omega \big((\ln n)^3\big)$, $\textrm{Var}[X] =
o\mathlarger{\mathlarger{\big(}}\frac{\{\mathbb{E}[X]\}^2}{ n(\ln
n)^2 }\mathlarger{\mathlarger{\big)}}$ and $|\alpha_n| = o(\ln n)$,
then
\begin{align}
 & \lim_{n \to \infty}\mathbb{P} \big[\hspace{2pt}\textrm{Graph }G(n,P_n,\mathcal {D})\textrm{
is $k$-robust}.\hspace{2pt}\big] \nonumber \\
& \quad =
\begin{cases} 0, &\textrm{ if $\lim_{n \to \infty}{\alpha_n}
=-\infty$}, \\  1, &\textrm{ if $\lim_{n \to \infty}{\alpha_n}
=\infty$.} \end{cases} \nonumber
 \end{align}

\end{thm}\qedb

\subsubsection{$k$-Robustness in Binomial Random Intersection
Graphs}

Theorem \ref{thm:rig:rb} below gives a zero--one law for asymptotic
$k$-robustness in a binomial random intersection graph.

\begin{thm} \label{thm:rig:rb} For a binomial random intersection graph
$G_b(n,P_n,p_n)$, with a sequence $\alpha_n$ for all $n $ defined
through
\begin{align}
  {p_n}^2 P_n & =
 \frac{\ln  n + {(k-1)} \ln \ln n + {\alpha_n}}{n},  \label{thm:rig:pe:rb}
\end{align}
if $P_n = \Omega \big(n(\ln n)^5\big)$ and $|\alpha_n| = o(\ln n)$,
then
\begin{align}
 & \lim_{n \to \infty}\mathbb{P} \big[\hspace{2pt}\textrm{Graph }G_b(n,P_n,p_n)\textrm{
is $k$-robust}.\hspace{2pt}\big] \nonumber \\
& \quad =
\begin{cases} 0, &\textrm{ if $\lim_{n \to \infty}{\alpha_n}
=-\infty$}, \\  1, &\textrm{ if $\lim_{n \to \infty}{\alpha_n}
=\infty$.} \end{cases} \nonumber
 \end{align}

 \end{thm}\qedb

\subsubsection{$k$-Robustness in Uniform Random Intersection Graphs}

 Theorem \ref{thm:urig:rb} below gives a zero--one law
for asymptotic $k$-robustness in a uniform random intersection
graph.

\begin{thm} \label{thm:urig:rb} For a uniform random intersection graph
$G_u(n,P_n,K_n)$, with a sequence $\alpha_n$ for all $n $ defined
through
\begin{align}
 \frac{{K_n}^2}{P_n} & = \frac{\ln  n + {(k-1)} \ln \ln n +
 {\alpha_n}}{n}, \label{thm:urig:pe:rb}
\end{align}
if $K_n = \Omega \big((\ln n)^3\big)$ and $|\alpha_n| = o(\ln n)$,
then
\begin{align}
 & \lim_{n \to \infty}\mathbb{P} \big[\hspace{2pt}\textrm{Graph }G_u(n,P_n,K_n)\textrm{
is $k$-robust}.\hspace{2pt}\big] \nonumber \\
& \quad =
\begin{cases} 0, &\textrm{ if $\lim_{n \to \infty}{\alpha_n}
=-\infty$}, \\  1, &\textrm{ if $\lim_{n \to \infty}{\alpha_n}
=\infty$.} \end{cases} \nonumber
 \end{align}

\end{thm} \qedb

In view of Theorems \ref{thm:grig}--\ref{thm:urig:rb}, for each
general/binomial/uniform random intersection graph, its
$k$-connectivity and $k$-robustness asymptotically obey the same
zero--one laws. Moreover, these zero--one laws are all \emph{sharp} since
$|\alpha_n|$ can be much smaller compared to $\ln n$; e.g., even
$\alpha_n = \pm c \cdot \ln \ln \cdot
\cdot \cdot \ln n$ with an arbitrary positive constant $c$ satisfies $\lim_{n \to
\infty}{\alpha_n} =\pm\infty$.

%

%
%

\section{Auxiliary Facts and Lemmas} \label{sec:factlem}


%
%
%
%


We present a few facts and lemmas which are used to establish the
theorems. To begin with, recalling that $k$ does not scale
 with $n$, we obtain Facts \ref{fact1} and \ref{fact_ln_n_n} below, whose proofs are
 straightforward and thus omitted here.\vspace{1pt}

 \begin{fact} \label{fact1}

For $|\alpha_n| = o(\ln n)$, it holds that
\begin{align}
& \frac{\ln  n + {(k-1)} \ln \ln n + {\alpha_n}}{n}
 \sim \frac{\ln  n}{n}. \nonumber 
 \end{align}
\end{fact}\qedb

\begin{figure*}[t]
\begin{center}
  \begin{tikzpicture}
  [scale=.9,auto=left]
  \node (v1) at (1,3) {Theorem 3};
  \node (v2) at (6,3) {Theorem 2};
  \node (v3) at (6,3.5) {Lemma 1};
  \node (v4) at (6,2.5) {Lemma 3};
   \node (v5) at (10,3) {Theorem 5};
   \node (l52) at (10,2.5) {Lemma 5};
   \node (l5) at (10.7,2.48) {~};
   \node (v6) at (14,3) {Theorem 6};
    \node (v56) at (12,3.038){~} ;
    \node (v567) at (11.991,2.962){~} ;
     \node (v25) at (8,3){~} ;
     \node (v251) at (8,2.962){~} ;
     \node (v252) at (8,3.038){~} ;
    \node (v7) at (17.5,2.1) {Theorem 4};
     \node (v8) at (1,1.2) {Lemma 2};
     \node (v82) at (1.65,1.13) {~};
     \node (v92) at (3.5,2.5) {Lemma 4};
      \node (v9) at (3.4,2.55) {~};
     \node (v10) at (4.53,3.063) {~};
     \node (v11) at (2.655,1.827) {~};
     \node (v111) at (2.95,1.735) {~};
     \node (v112) at (2.948,1.873) {~};
     \node (v12) at (2.25,1.5) {~};
     \node (v13) at (4.427,1.827) {Theorem 1};
\node (v46) at (15.687,2.13){~} ;
\node (v461) at (16,2.063){~} ;
\node (v462) at (16,2.139){~} ;
     \draw[->] (v1) -- (v2);
     \draw[->] (v2) -- (v5);
      \draw[->] (v5) -- (v6);
       \draw (v6) -- (v461);
        \draw(v9) -- (v10);
        \draw(v1) -- (v111);
         \draw(v8) -- (v112);
         \draw[->] (v11) -- (v13);
         \draw (v1) edge[out=16,in=164] (v567);
          \draw (v3) -- (v251);
          \draw(v4) -- (v252);
          \draw(l52) -- (v56);
          \draw(v82) -- (v462);
        \draw[->] (v46) -- (v7);

 \end{tikzpicture}
\caption{A plot illustrating the steps of deriving the theorems from the lemmas, with the arrows indicating the directions. For example, Theorem 3 and Lemma 2 are used to prove Theorem 1.\vspace{-20pt}} \label{fig:flow}
\end{center}
\end{figure*}
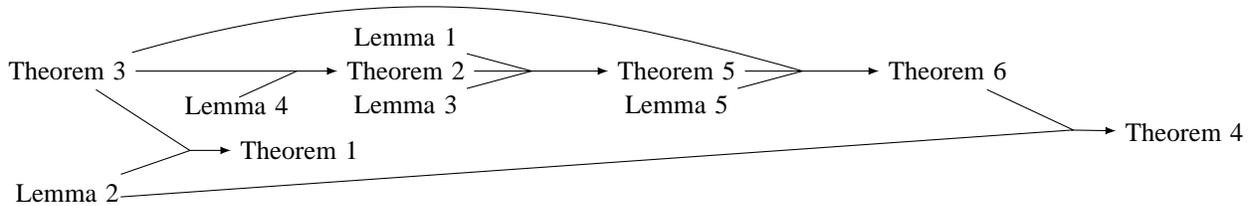

\begin{fact} \label{fact_ln_n_n}

For $|\alpha_n| = o(\ln n)$, we have
\begin{align}
& \frac{\ln  n + {(k-1)} \ln \ln n + {\alpha_n}}{n}
 \cdot \left[1 \pm
 o\left(\frac{1}{ \ln n}\right)\right] \nonumber \\
  & \quad = \frac{\ln  n +
{(k-1)} \ln \ln n + {\alpha_n} \pm o(1)}{n}, \nonumber 
 \end{align}
 and
 \begin{align}
& \frac{\ln  n + {(k-1)} \ln \ln n + {\alpha_n}}{n}
 \cdot \left[1 \pm
 O\left(\frac{1}{ \ln n}\right)\right] \nonumber \\
  & \quad = \frac{\ln  n +
{(k-1)} \ln \ln n + {\alpha_n} \pm O(1)}{n}. \nonumber 
 \end{align}
\end{fact}\qedb

Lemma \ref{er_robust} below presents the result on $k$-robustness
of an Erd\H{o}s-R\'enyi graph. An Erd\H{o}s--R\'{e}nyi graph
$G(n,\hat{p}_n)$ \cite{citeulike:4012374} is defined on a set of $n$
nodes such that any two nodes have an edge in between independently
with probability $\hat{p}_n$.\vspace{1pt}

\begin{lem}
\label{er_robust} For an Erd\H{o}s--R\'{e}nyi graph
$G(n,\hat{p}_n)$, with a sequence $\alpha_n$ for all $n$ through \vspace{-2pt}
\begin{align}
\hat{p}_n = & \frac{\ln  n + {(k-1)} \ln \ln n + {\alpha_n}}{n}
 \label{hat_pn},   \vspace{-2pt}
 \end{align}
then it holds that \vspace{-2pt}
\begin{align}
 & \lim_{n \to \infty} \hspace{-1pt}\mathbb{P} \big[G(n,\hat{p}_n)\textrm{
is $k$-robust}.\big]  = \begin{cases} 0, \textrm{ if $\lim_{n \to
\infty}{\alpha_n} \hspace{-2pt}=\hspace{-2pt}-\infty$}, \\  1,
\textrm{ if $\lim_{n \to \infty}{\alpha_n}
\hspace{-2pt}=\hspace{-2pt}\infty$.}
\end{cases} \vspace{-2pt} \label{er_rb}
 \end{align}
\end{lem}\qedb

To prove Lemma \ref{er_robust}, we note the following three
facts. (a) The desired result
(\ref{er_rb}) with $|\alpha_n| = o(\ln \ln n)$ is demonstrated in
\cite[Theorem 3]{6425841}. (b) By \cite[Facts 3 and 7]{zz}, for any
monotone increasing graph property $\mathcal {I}$, the probability
that graph $G(n,\hat{p}_n)$ has property $\mathcal {I}$ is
non-decreasing as $\hat{p}_n$ increases. (c) $k$-robustness is a monotone increasing graph property
according to \cite[Lemma 3]{6481629}. In view of (a) (b) and (c)
above, we obtain Lemma \ref{er_robust}. 


Throughout Lemmas \ref{lem:cp}--\ref{cp_urig_rig} below, $\mathcal {I}$ is
an arbitrary monotone increasing graph property, where a graph
property is called monotone increasing if it holds under the
addition of edges. Except Lemma 4 which is from \cite[Lemma 4]{Rybarczyk}, the proofs of Lemmas 2, 3 and 5 are deferred to
 Section \ref{sec:prf:fact:lem}.




%

\begin{lem} \label{lem:cp}
Let $X$ be a random variable with \vspace{2pt}probability
distribution $\mathcal {D}$. If $\textrm{Var}[X] =
o\mathlarger{\mathlarger{\big(}}\frac{\{\mathbb{E}[X]\}^2}{n (\ln
n)^2 }\mathlarger{\mathlarger{\big)}}$, then there exists
$\epsilon_n = o\big(\frac{1}{\ln  n}\big)$ such that \vspace{-2pt}
\begin{align}
 & \mathbb{P} \big[\hspace{2pt}\textrm{Graph }G(n,P_n,\mathcal {D})\textrm{
has $\mathcal {I}$}.\hspace{2pt}\big] \nonumber \\
&  \geq \mathbb{P} \big[\hspace{2pt}\textrm{Graph }G_u(n, P_n,
 (1 - \epsilon_n)\mathbb{E}[X])\textrm{ has $\mathcal {I}$}.
  \hspace{2pt}\big]  - o(1), \vspace{-2pt}\nonumber
 \end{align}
and \vspace{-2pt}
\begin{align}
 & \mathbb{P} \big[\hspace{2pt}\textrm{Graph }G(n,P_n,\mathcal {D})\textrm{
has $\mathcal {I}$}.\hspace{2pt}\big] \nonumber \\
&  \leq \mathbb{P} \big[\hspace{2pt}\textrm{Graph }G_u(n, P_n,
 (1 + \epsilon_n)\mathbb{E}[X])\textrm{ has $\mathcal {I}$}.
  \hspace{2pt}\big]  + o(1).\vspace{-2pt}\nonumber
 \end{align}

\end{lem}\qedb

\begin{lem} \label{cp_rig_er}

If $p_n = O\left( \frac{1}{n\ln n} \right)$\vspace{3pt} and ${p_n}^2
P_n =
 O\left( \frac{1}{\ln n} \right)$, then there exists $\hat{p}_n =
{p_n}^2 P_n \cdot \left[1-
 O\left(\frac{1}{ \ln n}\right)\right]$ such that  
\begin{align}
 & \mathbb{P}[\hspace{2pt}\textrm{Graph }G_b(n,P_n,p_n)
  \textrm{ has }\mathcal {I}.
\hspace{2pt}] \nonumber
\\ & \quad \geq
 \mathbb{P}[\hspace{2pt}\textrm{Graph }G(n,\hat{p}_n)
 \textrm{ has }\mathcal {I}.\hspace{2pt} ] - o(1).
\label{cp_res_rig_er}
 \end{align}

\end{lem}\qedb

\begin{lem}[{\hspace{-0.2pt}\cite[Lemma 4]{Rybarczyk}}] \label{rkgikg} If $p_n P_n = \omega\left( \ln n \right)$, and for all $n$
sufficiently large,
\begin{align}
K_{n,-}  & \leq p_n P_n - \sqrt{3(p_n P_n + \ln n) \ln n} ,
\nonumber \\
K_{n,+}  & \geq p_n P_n + \sqrt{3(p_n P_n + \ln n) \ln n}  ,
\nonumber
\end{align}
then
\begin{align}
 & \mathbb{P}[\hspace{2pt}\textrm{Graph }G_u(n,P_n,K_{n,-})
  \textrm{ has }\mathcal {I}.
\hspace{2pt}] - o(1)\nonumber
\\ &\quad \leq  \mathbb{P}[\hspace{2pt}\textrm{Graph }G_b(n,P_n,p_n)
 \textrm{ has }\mathcal {I}.\hspace{2pt} ]\nonumber
\\ & \quad\leq \mathbb{P}[\hspace{2pt}\textrm{Graph }G_u(n,P_n,K_{n,+})
  \textrm{ has }\mathcal {I}.
\hspace{2pt}]+ o(1) .  \nonumber 
 \end{align}
\end{lem} \vspace{2pt}\qedb

\begin{lem} \label{cp_urig_rig}

If $K_n \hspace{-1pt}=\hspace{-1pt} \omega\left( \ln n \right)$ and
$p_n \hspace{-1pt}=\hspace{-1pt} \frac{K_n}{P_n}
 \left(1 - \sqrt{\frac{3\ln
n}{K_n }}\hspace{2pt}\right)$, then   
\begin{align}
 & \mathbb{P}[\hspace{2pt}\textrm{Graph }G_u(n,P_n,K_n)
  \textrm{ has }\mathcal {I}.
\hspace{2pt}] \nonumber
\\ & \quad \geq
 \mathbb{P}[\hspace{2pt}\textrm{Graph }G_b(n,P_n,p_n)
 \textrm{ has }\mathcal {I}.\hspace{2pt} ] - o(1). \nonumber
 \end{align}

\end{lem}\qedb

Figure \ref{fig:flow} on the next page illustrates the steps of using the lemmas to prove the theorems. Note that the facts used in deriving the theorems are not shown in the plot for brevity.

\section{Establishing Theorems \ref{thm:grig}--\ref{thm:urig}}
\label{sec:thmprf:kcon}

Theorems \ref{thm:grig}--\ref{thm:urig} describe results on
$k$-connectivity for various random intersection graphs.  



\subsection{The Proof of Theorem \ref{thm:grig}}

We demonstrate Theorem \ref{thm:grig} with the help of Theorem
\ref{thm:urig}, the proof of which is detailed in Section
\ref{prf:thm:urig}.

For any $\epsilon_n = o\left(\frac{1}{\ln  n}\right)$, it is clear
that
\begin{align}
 (1\pm \epsilon_n)^2 & = 1\pm 2\epsilon_n + {\epsilon_n}^2
  =  1\pm  o\left(\frac{1}{\ln  n}\right). \label{eps_2}
 \end{align}
We recall conditions (\ref{thm:grig:pe}) and $|\alpha_n| = o(\ln
n)$, which together with (\ref{eps_2}) and Fact \ref{fact_ln_n_n}
yields
\begin{align}
 \frac{\big\{(1\pm \epsilon_n)\mathbb{E}[X]\big\}^2}{P_n}
   & = \frac{\ln  n + {(k-1)}
 \ln \ln n + {\alpha_n} \pm o(1)}{n}.\label{pe_epsilon4}
 \end{align}
With $\mathbb{E}[X] = \Omega \big(\sqrt{\ln n}\hspace{2pt}\big)$ and
$\epsilon_n = o\left(\frac{1}{\ln n}\right)$, it follows that $(1 \pm \epsilon_n)\mathbb{E}[X]
 =  \Omega \big(\sqrt{\ln n}\hspace{2pt}\big)$, which along with (\ref{pe_epsilon4}) and $|\alpha_n| = o(\ln n)$ enables the use of Theorem \ref{thm:urig} to derive
\begin{align}
 & \lim_{n \to \infty}\mathbb{P} \big[ G_u(n, P_n,
 (1 \pm \epsilon_n)\mathbb{E}[X])
 \textrm{
is $k$-connected}. \big] \nonumber  \\
&   =
\begin{cases} 0, &\textrm{ if $\lim_{n \to \infty}{\alpha_n}
=-\infty$}, \\  1, &\textrm{ if $\lim_{n \to \infty}{\alpha_n}
=\infty$,} \\ e^{- \frac{e^{-\alpha ^*}}{(k-1)!}},
 &\textrm{ if $\lim_{n \to \infty}{\alpha_n}
=\alpha ^* \in (-\infty, \infty)$.} \end{cases} \label{kconn}
 \end{align}
Since $k$-connectivity is a monotone increasing graph property
\cite{zz}, Theorem \ref{thm:grig} is proved by (\ref{kconn}) and
Lemma \ref{lem:cp}. \qeda

\subsection{The Proof of Theorem \ref{thm:rig}} \label{pfrig}

From Lemma \ref{rkgikg} and Theorem \ref{thm:urig}, the proof of
Theorem \ref{thm:rig} is completed once we show that with
$K_{n,\pm}$ defined by
\begin{align}
K_{n,\pm}  & = p_n P_n \pm \sqrt{3(p_n P_n + \ln n) \ln n} ,
\label{Kngeqsb}
\end{align}
under conditions of Theorem \ref{thm:rig}, we have $K_{n,\pm}    =
\Omega \big(\sqrt{\ln n}\hspace{2pt}\big)$ and with $\alpha_{n,\pm}$
defined by
\begin{align}
 \frac{{K_{n,\pm}}^2}{P_n} & = \frac{\ln  n + {(k-1)} \ln \ln n +
 {\alpha_{n,\pm}}}{n},  \label{thm:urig:peab}
 \end{align}
then
\begin{align}
\alpha_{n,\pm}  &   = \alpha_{n} \pm o(1).  \label{thm:urig:peaaph}
\end{align}

From conditions (\ref{thm:rig:pe}) and $|\alpha_n| = o(\ln n)$, and
Fact \ref{fact1}, it is clear that
\begin{align}
 {p_n}^2 P_n & \sim
 \frac{\ln  n}{n} .\label{thm:rig:pe_sim}
\end{align}
Substituting (\ref{thm:rig:pe_sim}) and condition $P_n = \omega
\big(n(\ln n)^5\big)$ into (\ref{Kngeqsb}), it holds that
\begin{align}
K_{n,\pm}  &  = \omega \big((\ln n)^3\big) = \Omega \big(\sqrt{\ln
n}\hspace{2pt}\big) , \label{zh1}
\end{align}
and
\begin{align}
 \frac{{K_{n,\pm}}^2}{P_n} & =   {p_n}^2 P_n \cdot \bigg[1 \pm o\bigg(\frac{1}{\ln n}\bigg)\bigg]. \label{thm:urig:pea}
\end{align}
Then from (\ref{thm:rig:pe}) (\ref{thm:urig:peab})
(\ref{thm:urig:pea}) and Fact \ref{fact_ln_n_n}, we obtain
(\ref{thm:urig:peaaph}). As explained before, with
(\ref{thm:urig:peab}) (\ref{thm:urig:peaaph}) and (\ref{zh1}),
Theorem \ref{thm:rig} is proved from Lemma \ref{rkgikg} and Theorem
\ref{thm:urig}. \qeda

As noted in Remark \ref{rm}, to prove
only the zero--one law but not the asymptotically exact probability
result in Theorem \ref{thm:rig}, condition $P_n = \omega
\big(n(\ln n)^5\big)$ can be weakened as $P_n = \Omega \big(n(\ln
n)^5\big)$. This can be seen by the argument that under $P_n =
\Omega \big(n(\ln n)^5\big)$, (\ref{thm:urig:peaaph}) can be
weakened as $\alpha_{n,\pm}   = \alpha_{n} \pm O(1)$, which can
still used to establish the zero--one law.

\subsection{The Proof of Theorem \ref{thm:urig}}
\label{prf:thm:urig}

We derive in \cite{mobihocQ1} the asymptotically exact probability
and an asymptotic zero--one law for $k$-connectivity in graph
$G(n,\tilde{p}_n)\bcap G_u(n,P_n,K_n)$, which is the superposition
of an Erd\H{o}s--R\'{e}nyi graph $G(n,\tilde{p}_n)$ on a uniform
random intersection graph $G_u(n,P_n,K_n)$. Setting $\tilde{p}_n =
1$, graph
$G(n,\tilde{p}_n)\bcap G_u(n,P_n,K_n)$ becomes $ G_u(n,P_n,K_n)$. Then with
$\tilde{p}_n = 1$, we obtain from \cite[Theorem 1]{mobihocQ1} that
if $P_n = \Omega (n)$ and
\begin{align}
1- \binom{P_n- K_n}{K_n} \bigg/ \binom{P_n}{K_n} & = \frac{\ln  n +
{(k-1)} \ln \ln n + {\beta_n}}{n}, \label{PnKnKn}
\end{align}
then
\begin{align}
 & \lim_{n \to \infty}\mathbb{P} \left[\hspace{2pt}G_u(n,P_n,K_n)\textrm{
is $k$-connected}.\hspace{2pt}\right] \nonumber \\
& ~~  =
\begin{cases} 0, &\textrm{ if $\lim_{n \to \infty}{\beta_n}
=-\infty$},  \\  1, &\textrm{ if $\lim_{n \to \infty}{\beta_n}
=\infty$,}  \\ e^{- \frac{e^{-\beta ^*}}{(k-1)!}},
 &\textrm{ if $\lim_{n \to \infty}{\beta_n}
=\beta ^* \in (-\infty, \infty)$.} \end{cases} \label{beta}
 \end{align}
Note that if $\beta_n = \alpha_n \pm o(1)$, then (i) $\lim_{n \to
\infty}{\beta_n}$ exists if and only if $\lim_{n \to
\infty}{\alpha_n}$ exists; and (ii) when they both exist, $\lim_{n
\to \infty}{\beta_n} = \lim_{n \to \infty}{\alpha_n}$. Therefore,
Theorem \ref{thm:urig} is proved\vspace{1pt} once we show $P_n =
\Omega (n)$ and (\ref{PnKnKn}) with $\beta_n = \alpha_n \pm o(1)$
given conditions $K_n = \Omega \big(\sqrt{\ln n}\hspace{2pt}\big)$,
$|\alpha_n| = o(\ln n)$ and (\ref{thm:urig:pe}).

 From $|\alpha_n| = o(\ln n)$, (\ref{thm:urig:pe}) and Fact
\ref{fact1}, it holds that
\begin{align}
 \frac{{K_n}^2}{P_n} & \sim \frac{\ln n}{n} , \label{KKP}
\end{align}
which along with $K_n = \Omega \big(\sqrt{\ln n}\hspace{2pt}\big)$
yields
\begin{align}
P_n  & \sim \frac{n{K_n}^2}{\ln n} = \Omega (n).\nonumber
\end{align}

We derive in \cite[Lemma 8]{ZhaoYaganGligor} that
\begin{align}
1\hspace{-.5pt}-\hspace{-.5pt} \binom{P_n- K_n}{K_n}
\hspace{-.5pt}\bigg/\hspace{-.5pt} \binom{P_n}{K_n} &
\hspace{-.25pt}=\hspace{-.25pt} \frac{{K_n}^2}{P_n}
\hspace{-.25pt}\cdot\hspace{-.25pt} \bigg[1 \pm O\bigg(
\frac{{K_n}^2}{P_n} \bigg) \bigg]. \label{PnKK}
\end{align}
Applying (\ref{KKP}) to (\ref{PnKK}),
\begin{align}
1\hspace{-.5pt}- \hspace{-.5pt}\binom{P_n- K_n}{K_n}
\hspace{-.5pt}\bigg/\hspace{-.5pt} \binom{P_n}{K_n} &\hspace{-.25pt}
= \hspace{-.25pt}\frac{{K_n}^2}{P_n}
\hspace{-.25pt}\cdot\hspace{-.25pt} \bigg[1 \pm o\left(\frac{1}{ \ln
n}\right)\bigg], \nonumber
\end{align}
which together with (\ref{thm:urig:pe}) and Fact \ref{fact_ln_n_n}
leads to (\ref{PnKnKn}) with condition $\beta_n = \alpha_n \pm
o(1)$. Since we have proved $P_n = \Omega (n)$ and (\ref{PnKnKn})
with $\beta_n = \alpha_n \pm o(1)$, Theorem \ref{thm:urig} follows
from (\ref{beta}).

\qeda

\section{Establishing Theorems \ref{thm:grig:rb}--\ref{thm:urig:rb}}
\label{sec:thmprf:krb}

Theorems \ref{thm:grig:rb}--\ref{thm:urig:rb} present results on
$k$-robustness for various random intersection graphs.  

%

\subsection{The Proof of Theorem \ref{thm:grig:rb}} \label{prf:thm:grig:rb}

Similar to the process of proving Theorem \ref{thm:grig} with the
help of Theorem \ref{thm:urig}, we demonstrate Theorem
\ref{thm:grig:rb} using Theorem \ref{thm:urig:rb}, the proof of
which is given in Section \ref{prf:thm:urig:rb}.

 Note that condition (\ref{thm:grig:pe:rb}) is the same as
(\ref{thm:grig:pe}), and condition $|\alpha_n| = o(\ln n)$ holds.
Then as shown in Theorem \ref{thm:grig}, for any $\epsilon_n =
o\left(\frac{1}{\ln n}\right)$, from (\ref{thm:grig:pe})
(\ref{eps_2}), $|\alpha_n| = o(\ln n)$ and Fact
\ref{fact_ln_n_n},\vspace{2pt} we obtain (\ref{pe_epsilon4}) here.
From $\mathbb{E}[X] = \Omega \big((\ln n)^3\big)$ and $\epsilon_n =
o\left(\frac{1}{\ln n}\right)$, it follows that $ (1 \pm \epsilon_n)\mathbb{E}[X] = \Omega \big((\ln n)^3\big) $, which along with (\ref{pe_epsilon4}) enables the use of
Theorem
\ref{thm:urig:rb} to yield that for $\mathbb{E}[X] = \Omega
\big((\ln n)^3\big)$ and any $\epsilon_n = o\big(\frac{1}{\ln
n}\big)$, we have
\begin{align}
 & \lim_{n \to \infty}\mathbb{P} \big[ G_u(n, P_n,
 (1 \pm \epsilon_n)\mathbb{E}[X])
 \textrm{
is $k$-robust}. \big] \nonumber  \\
&   =
\begin{cases} 0, &\textrm{ if $\lim_{n \to \infty}{\alpha_n}
=-\infty$}, \\  1, &\textrm{ if $\lim_{n \to \infty}{\alpha_n}
=\infty$.} \end{cases} \label{kconn:rb}
 \end{align}
Since $k$-robustness is a monotone increasing graph property
according to \cite[Lemma 3]{6481629}, Theorem \ref{thm:grig:rb} is
proved by (\ref{kconn:rb}) and Lemma \ref{lem:cp}. \qeda

\subsection{The Proof of Theorem \ref{thm:rig:rb}} \label{prf:thm:rig:rb}

Since $k$-robustness implies $k$-connectivity by \cite[Lemma
1]{6425841}, the zero law of Theorem \ref{thm:rig:rb} is clear from
 Theorem \ref{thm:rig} and Remark \ref{rm} in view that under conditions of Theorem \ref{thm:rig:rb},
  if $\lim_{n \to \infty}{\alpha_n}
=-\infty$, \begin{align}
 & \mathbb{P} \big[\hspace{2pt}G_b(n,P_n,p_n)\textrm{
is $k$-robust}.\hspace{2pt}\big] \nonumber \\
& \leq \mathbb{P} \big[\hspace{2pt}G_b(n,P_n,p_n)\textrm{ is
$k$-connected}.\hspace{2pt}\big] \to 0,\textrm{ as }n \to \infty.
\label{prf:thm:rig:rb1}
 \end{align}

Below we prove the one law of Theorem \ref{thm:rig:rb}.  
 Note that (\ref{thm:rig:pe:rb}) is the same as (\ref{thm:rig:pe}),
and we have condition $|\alpha_n| = o(\ln n)$. Then as proved in
Theorem \ref{thm:rig}, given (\ref{thm:rig:pe}) and $|\alpha_n| =
o(\ln n)$, we obtain (\ref{thm:rig:pe_sim}), which together with
condition $P_n = \Omega \big(n(\ln n)^5\big)$ leads to
\begin{align}
p_n & \sim
 \sqrt{\frac{\ln  n}{nP_n}} = O\Bigg(\sqrt{\frac{\ln  n}{n^2(\ln n)^5}}\hspace{2pt}\Bigg)
 = O\bigg(\frac{1}{n(\ln n)^2}\bigg).   \label{thm:rig:pnx}
\end{align}
Noting that (\ref{thm:rig:pnx}) implies condition $p_n = O\left( \frac{1}{n\ln n} \right)$ in Lemma \ref{cp_rig_er}, we apply Lemmas \ref{er_robust} and \ref{cp_rig_er}, and
condition (\ref{thm:rig:pe:rb}) to derive the following: there exists
$\hat{p}_n = \frac{\ln  n + {(k-1)} \ln \ln n + {\alpha_n} -
O(1)}{n}$ such that if $\lim_{n \to \infty}{\alpha_n} = \infty$,
\begin{align}
 & \mathbb{P}[\hspace{2pt}\textrm{Graph }G_b(n,P_n,p_n)
  \textrm{ is $k$-robust.}
\hspace{2pt}] \nonumber
\\ &  \geq \hspace{-1pt}
 \mathbb{P}[\hspace{1.5pt}\textrm{Graph }G(n,\hat{p}_n)
 \textrm{ is $k$-robust.}\hspace{1.5pt} ]  \hspace{-1pt}- \hspace{-1pt} o(1)
  \hspace{-1pt}\to \hspace{-1pt} 1,\textrm{ as }n  \hspace{-1pt}\to \hspace{-1pt} \infty. \label{prf:thm:rig:rb2}
 \end{align}
The proof of Theorem \ref{thm:rig:rb} is completed via
(\ref{prf:thm:rig:rb1}) and (\ref{prf:thm:rig:rb2}).  \qeda

\subsection{The Proof of Theorem \ref{thm:urig:rb}} \label{prf:thm:urig:rb}

The zero law of Theorem \ref{thm:urig:rb} is proved below by an
 approach similar to that of Theorem \ref{thm:rig:rb}. Since
$k$-robustness implies $k$-connectivity by \cite[Lemma 1]{6425841},
the zero law of Theorem \ref{thm:urig:rb} is clear from
 Theorem \ref{thm:urig} in view that under conditions of Theorem \ref{thm:urig:rb}, if $\lim_{n \to \infty}{\alpha_n}
=-\infty$,
\begin{align}
 & \mathbb{P} \big[\hspace{2pt}G_u(n,P_n,K_n)\textrm{
is $k$-robust}.\hspace{2pt}\big] \nonumber \\
& \leq \mathbb{P} \big[\hspace{2pt}G_u(n,P_n,K_n)\textrm{ is
$k$-connected}.\hspace{2pt}\big] \to 0,\textrm{ as }n \to \infty.
\label{prf:thm:urig:rb1}
 \end{align}

Below we establish the one law of Theorem \ref{thm:urig:rb} with the
help of Theorem \ref{thm:rig:rb}. Given $K_n = \Omega \big((\ln
n)^3\big) = \omega\left( \ln n \right)$, we use Lemma
\ref{cp_urig_rig} to obtain that with $p_n$ set by
\begin{align}
p_n & =  \frac{K_n}{P_n}
 \left(1 - \sqrt{\frac{3\ln
n}{K_n }}\hspace{2pt}\right), \label{pnexpr}
 \end{align}
it holds that
\begin{align}
 & \mathbb{P}[\hspace{2pt}\textrm{Graph }G_u(n,P_n,K_n)
  \textrm{ is $k$-robust.}
\hspace{2pt}] \nonumber
\\ & \quad \geq
 \mathbb{P}[\hspace{2pt}\textrm{Graph }G_b(n,P_n,p_n)
 \textrm{ is $k$-robust.}\hspace{2pt} ] - o(1). \label{robustcomp}
 \end{align}

Note that (\ref{thm:urig:pe:rb}) is the same as (\ref{thm:urig:pe});
and $|\alpha_n| = o(\ln n)$ holds as a condition. Then as shown in
Theorem \ref{thm:urig}, from (\ref{thm:urig:pe}), $|\alpha_n| =
o(\ln n)$
and Fact \ref{fact_ln_n_n}, we obtain (\ref{KKP}) here, %
%
 which together with $K_n = \Omega \big((\ln n)^3\big) $ results in
\begin{align}
P_n & \sim \frac{n{K_n}^2}{\ln n} = \Omega \big(n(\ln n)^5\big),
\label{Pnlnn5}
\end{align}
From $K_n = \Omega \big((\ln n)^3\big) $ and (\ref{pnexpr}), it
follows that
\begin{align}
{p_n}^2 P_n & =  \left[\frac{K_n}{P_n}
 \left(1 - \sqrt{\frac{3\ln
n}{K_n }}\hspace{2pt}\right)\right]^2 \cdot P_n \nonumber \\ & =
\frac{{K_n}^2}{P_n}
 \cdot \left[1 -
 O\left(\frac{1}{ \ln n}\right)\right] .\label{pnPnlnn}
 \end{align}
By (\ref{thm:urig:pe:rb}) (\ref{pnPnlnn}) and Fact
\ref{fact_ln_n_n}, it is clear that
\begin{align}
  {p_n}^2 P_n & =
 \frac{\ln  n + {(k-1)} \ln \ln n + {\alpha_n} - O(1)}{n}. \label{thm:rig:pe:rbcdx}
\end{align}
Given (\ref{Pnlnn5}) (\ref{thm:rig:pe:rbcdx}) and $|\alpha_n| =
o(\ln n)$, we use Theorem \ref{thm:rig:rb} and (\ref{robustcomp}) to
get that if $\lim_{n \to \infty}{\alpha_n} =\infty$,
\begin{align}
 & \mathbb{P} \big[\hspace{2pt}G_u(n,P_n,K_n)\textrm{
is $k$-robust}.\hspace{2pt}\big] \to 1,\textrm{ as }n \to \infty.
\label{prf:thm:urig:rb2}
 \end{align}
The proof of Theorem \ref{thm:urig:rb} is completed via
(\ref{prf:thm:urig:rb1}) and (\ref{prf:thm:urig:rb2}).  \qeda



\section{Establishing Lemmas in Section \ref{sec:factlem}} \label{sec:prf:fact:lem}

Lemmas \ref{er_robust} and \ref{rkgikg} are clear in Section \ref{sec:factlem}. Below we prove Lemmas \ref{lem:cp}, \ref{cp_rig_er} and \ref{cp_urig_rig}.

 \subsection{The Proof of Lemma \ref{lem:cp}}

According to \cite[Lemma 3]{Rybarczyk}, for any monotone increasing
graph property $\mathcal {I}$ and any $|\epsilon_n|<1$,
\begin{align}
 & \mathbb{P} \big[ G(n,P_n,\mathcal {D})\textrm{
has $\mathcal {I}$}. \big]   -  \mathbb{P} \hspace{-1pt} \big[
G_u(n,\hspace{-1.3pt}P_n,\hspace{-1.3pt}(1\hspace{-1.3pt}-\hspace{-1.3pt}
\epsilon_n)\mathbb{E}[X])\textrm{\hspace{-.2pt} has $\mathcal
{I}$}.\hspace{-.2pt} \big] \nonumber \\
&  \geq  \big\{ 1 - \mathbb{P}\hspace{-.2pt}[X \hspace{-1.3pt} <
\hspace{-1.3pt}
(1\hspace{-1.3pt}-\hspace{-1.3pt}\epsilon_n)\mathbb{E}[X]] \big\}^n
- 1, \label{coupling1}
 \end{align}
and
\begin{align}
 &  \mathbb{P} \big[ G(n,P_n,\mathcal {D})\textrm{
has $\mathcal {I}$}. \big]  - \mathbb{P} \hspace{-1pt} \big[
G_u(n,\hspace{-1.3pt}P_n,\hspace{-1.3pt}(1\hspace{-1.3pt}+\hspace{-1.3pt}
\epsilon_n)\mathbb{E}[X])\textrm{\hspace{-.2pt} has $\mathcal
{I}$}.\hspace{-.2pt} \big] \nonumber \\
&  \leq 1 -  \big\{1 - \mathbb{P}\hspace{-.2pt}[X \hspace{-1.3pt}
> \hspace{-1.3pt}
(1\hspace{-1.3pt}+\hspace{-1.3pt}\epsilon_n)\mathbb{E}[X]] \big\}^n.
\label{coupling2}
 \end{align}

 By (\ref{coupling1})
(\ref{coupling2}) and the fact that $\lim\limits_{n \to
\infty}(1-m_n)^n = 1$ for $m_n = o\big(\frac{1}{n}\big)$ (this can
be proved by a simple Taylor series expansion as in \cite[Fact
2]{ZhaoYaganGligor}), the proof of Lemma \ref{lem:cp} is
completed\vspace{2pt} once we demonstrate that with $\textrm{Var}[X]
= o\Big(\frac{\{\mathbb{E}[X]\}^2}{ n(\ln n)^2 }\Big)$, there exists
$\epsilon_n = o\big(\frac{1}{\ln n}\big)$ such that
\begin{align}
\bP{X <
 (1 - \epsilon_n)\mathbb{E}[X]} & = o\bigg(\frac{1}{n}\bigg) , \label{leq1esp_lem}
 \end{align}
 and
 \begin{align}
\bP{X >
 (1 + \epsilon_n)\mathbb{E}[X]} & = o\bigg(\frac{1}{n}\bigg). \label{geq1esp_lem}
 \end{align}
To prove (\ref{leq1esp_lem}) and (\ref{geq1esp_lem}),
Chebyshev's inequality yields
\begin{align}
\mathbb{P} \big[\hspace{2pt} |X-\mathbb{E}[X]| >
 \epsilon_n \mathbb{E}[X]\big]  & \leq
 \frac{\textrm{Var}[X]}{\big\{\epsilon_n \mathbb{E}[X]\big\}^2}.
  \label{thm:grig:Xbound_lem}
 \end{align}
We set $\epsilon_n$ by $\epsilon_n =
\sqrt[4]{\frac{n\textrm{Var}[X]}{\big\{\mathbb{E}[X]\big\}^2}} \cdot
\frac{1}{\sqrt{\ln n}} $.
 Then given condition $\textrm{Var}[X] = o\Big(\frac{\{\mathbb{E}[X]\}^2}{ n(\ln
n)^2 }\Big)$, we obtain
\begin{align}
 \epsilon_n & =
 o\Bigg(
  \sqrt[4]{\frac{1}{(\ln n)^2}} \hspace{2pt}\Bigg) \cdot \frac{1}{\sqrt{\ln n}}
   = o\Big(\frac{1}{\ln
n}\Big), \label{eps_lem}
\end{align}
and
\begin{align}
\frac{\textrm{Var}[X]}{\big\{\epsilon_n \mathbb{E}[X]\big\}^2}  & =
\sqrt{\frac{\textrm{Var}[X]}{n\big\{\mathbb{E}[X]\big\}^2}} \cdot
\ln n
  = o\bigg(\frac{1}{n}\bigg).
\label{varX_lem}
 \end{align}
By (\ref{thm:grig:Xbound_lem}) (\ref{eps_lem}) and (\ref{varX_lem}),
it is straightforward to see that (\ref{leq1esp_lem}) and
(\ref{geq1esp_lem}) hold with $\epsilon_n = o\big(\frac{1}{\ln
n}\big)$. Therefore, we have completed the proof of Lemma
\ref{lem:cp}. \qeda

\subsection{The Proof of Lemma \ref{cp_rig_er}}

In view of \cite[Lemma 3]{zz}, if ${p_n}^2 P_n <   1$ and
$p_n=o\left(\frac{1}{ n}\right)$, with $\hat{p}_n: = {p_n}^2 P_n \cdot \left(1 - n{p_n} + 2 {p_n} -
\frac{{p_n}^2 P_n}{2} \right)$,
then (\ref{cp_res_rig_er}) follows. Given conditions $p_n = O\left(
\frac{1}{n\ln n} \right)$ \vspace{2pt}and ${p_n}^2 P_n =
 O\left( \frac{1}{\ln n} \right)$ in Lemma \ref{cp_rig_er}, ${p_n}^2 P_n <   1$ and
\vspace{2pt}$p_n=o\left(\frac{1}{ n}\right)$ clearly hold. Then
Lemma \ref{cp_rig_er} is proved once we show $\hat{p}_n$ satisfies $\hat{p}_n = {p_n}^2 P_n \cdot
\left[1-
 O\left(\frac{1}{ \ln n}\right)\right]$, which is easy to see via
 \begin{align}
&  - n{p_n} + 2 {p_n} - \frac{{p_n}^2 P_n}{2} \nonumber \\
& \hspace{1pt} = \hspace{-1pt} (-n\hspace{-1pt}+\hspace{-1pt}2)
 \hspace{-1pt}\cdot\hspace{-1pt} O\left(
\frac{1}{n\ln n} \right) \hspace{-1pt}-\hspace{-1pt} \frac{1}{2}
\hspace{-1pt}\cdot\hspace{-1pt} O\left( \frac{1}{\ln n} \right)
\hspace{-1pt}=\hspace{-1pt} -
 O\left(\frac{1}{ \ln n}\right) . \nonumber
\end{align}
Hence, the proof of Lemma \ref{cp_rig_er} is completed. \qeda

\subsection{The Proof of Lemma \ref{cp_urig_rig}}

We use Lemma \ref{rkgikg} to prove Lemma \ref{cp_urig_rig}. From
conditions\vspace{1pt} $K_n = \omega\left( \ln n \right)$ and $p_n =
\frac{K_n}{P_n}
 \left(1 - \sqrt{\frac{3\ln
n}{K_n }}\hspace{2pt}\right)$, we first obtain $p_n P_n  = \omega\left( \ln n \right)$ and then
 for all $n$ sufficiently large,
\begin{align}
&  K_n - \left[ p_n P_n + \sqrt{3(p_n P_n + \ln n) \ln n}
\hspace{1.5pt}\right] \nonumber \\ & = K_n \sqrt{\frac{3\ln n}{K_n
}} - \sqrt{3\left[ K_n \left(1 - \sqrt{\frac{3\ln n}{K_n
}}\hspace{2pt}\right) + \ln n\right] \ln n} \nonumber
\\  & = \sqrt{3K_n\ln n}  -
\sqrt{3\left[K_n  \hspace{-1pt}+ \hspace{-1pt} \sqrt{\ln n} \left(
\sqrt{\ln n} \hspace{-1pt}- \hspace{-1pt}
\sqrt{3K_n}\hspace{2pt}\right) \right ] \hspace{-1pt} \ln n}
\nonumber \\  & \geq  0. \nonumber
\end{align}
Then by  Lemma \ref{rkgikg}, Lemma \ref{cp_urig_rig} is now
established. \qeda

\begin{figure}[!t]
  \centering
 \includegraphics[width=0.35\textwidth]{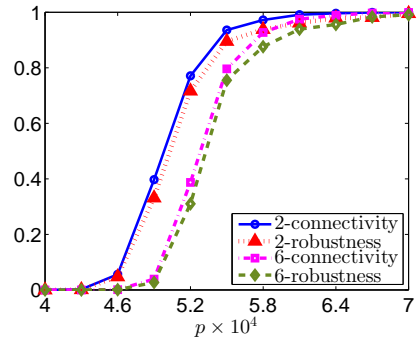} \vspace{-2pt}
\caption{A plot of the empirical probabilities that binomial random
intersection graph $G_b(n,P,p)$ has $k$-connectivity or
$k$-robustness as a function of $p$, with $n=2,000$, $P=20,000$ and
$k=2,6$. \vspace{-10pt} } \label{fig}
\end{figure}

\begin{figure}[!t]
  \centering
 \includegraphics[width=0.35\textwidth]{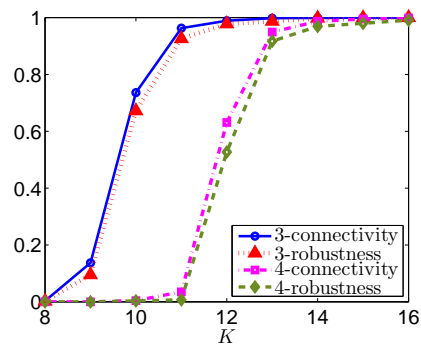} \vspace{-2pt}
 \caption{A plot of the empirical probabilities that uniform random intersection
graph $G_u(n,P,K)$ has $k$-connectivity or $k$-robustness as a
function of $K$, with $n=2,000$, $P=20,000$ and $k=3,4$.
\vspace{-10pt} }\label{figa}
\end{figure}

\section{Numerical Experiments} \label{sec:expe}

We present numerical experiments in the non-asymptotic regime to
confirm our theoretical results. 

Figure \ref{fig} depicts the probability that binomial random
intersection graph $G_b(n,P,p)$ has $k$-connectivity or
$k$-robustness, for $k = 2,6$. Similarly, Figure \ref{figa} illustrates the probability of $k$-connectivity or
$k$-robustness for
$k = 3, 4$ in uniform random intersection graph $G_u(n,P,K)$. In
all set of experiments, we fix the number of nodes at $n=2,000$ and
the object pool size $P = 20,000$. For each pair $(n,P,p)$ (resp.,
$(n,P,K)$), we generate $1,000$ independent samples of $G_b(n,P,p)$
(resp., $G_u(n,P,K)$) and count the number of times that the
obtained graphs are $k$-connected or $k$-robust. Then the counts
divided by $1,000$ become the corresponding empirical probabilities.
As illustrated in Figures \ref{fig} and \ref{figa}, there is an
evident threshold in the probabilities of $k$-connectivity
and $k$-robustness. Also, for each $k$, the curves of
$k$-connectivity and $k$-robustness are close to each other. These
numerical results are in agreement with our analytical findings in the theorems.

\section{Related Work} \label{related}

For connectivity (i.e., $k$-connectivity with $k=1$) in binomial
random intersection graph $G_b(n,P_n,p_n)$, Rybarczyk establishes
the exact probability \cite{2013arXiv1301.0466R} and a zero--one law
\cite{zz,2013arXiv1301.0466R}. 
 She
further shows a zero--one law for $k$-connectivity
\cite{zz,2013arXiv1301.0466R}. Our
Theorem \ref{thm:rig} provides not only a zero--one law,
 but also the exact probability to deliver a precise understanding of
$k$-connectivity.

 For connectivity in
uniform random intersection graph $G_u(n,P_n,K_n)$, Rybarczyk
\cite{ryb3} derives the exact probability and a zero--one law, while
Blackburn and Gerke \cite{r1}, Ya\u{g}an and Makowski
\cite{yagan}, and Zhao \emph{et al.} \cite{ISIT, ZhaoYaganGligor} also obtain zero--one laws. Rybarczyk \cite{zz}
implicitly shows a zero--one law for $k$-connectivity in
$G_u(n,P_n,K_n)$. Our Theorem \ref{thm:urig} also gives a zero--one law. In addition, it gives the exact probability to
provide an accurate understanding of $k$-connectivity.

For general random intersection graph $G(n,P_n,\mathcal {D})$,
Godehardt and Jaworski \cite{Models} investigate its degree
distribution and Bloznelis \emph{et al.} \cite{Rybarczyk} explore
its component evolution, but provides neither a zero--one law nor the
 exact probability of its $k$-connectivity property reported in our work.


To date, there have not been any results reported on the
($k$-)robustness of random intersection graphs by others. As noted
in Lemma \ref{er_robust}, Zhang and Sundaram \cite{6425841} present
a zero--one law for $k$-robustness in an Erd\H{o}s--R\'{e}nyi graph.
\vspace{-1pt} 

For random intersection graphs in this paper, two nodes have an edge in between if their object sets share at least one object. A natural variant is to define graphs with edges only between nodes which have at least $s$ objects in common (instead of just $1$) for some positive integer $s$. Zhao \emph{et al.} \cite{JZISIT14,QcompTech14,ANALCO} consider $k$-connectivity in graphs under this definition. In addition, ($k$)-connectivity of other random graphs have also been investigated in the literature \cite{ZhaoISIT2014,FJYGISIT2014}.

\section{Conclusion and Future Work}\vspace{-1pt}
\label{sec:Conclusion}

Under a general random intersection graph model, we derive sharp
zero--one laws for $k$-connectivity and $k$-robustness, as well as
the asymptotically exact probability of $k$-connectivity,
 where $k$ is an arbitrary positive integer. A future direction is to obtain the asymptotically exact probability of $k$-robustness for a precise characterization on the robustness strength. \vspace{-1pt}

   \section*{Acknowledgements} \vspace{-1pt}

 This research was supported in part
by CMU CyLab under the National Science Foundation grant
CCF-0424422 to the Berkeley TRUST STC. The views and conclusions contained in this document are those
of the authors and should not be interpreted as representing the official policies,
either expressed or implied, of any sponsoring institution, the
U.S. government or any other entity.

\end{document}